\documentclass[11pt]{iopart}

\usepackage{fullpage,iopams,amsthm,equation}

\newtheorem{theorem}{Theorem}

\newcommand{\R}{\mathbb R}

\begin{document}
\title{Capture-zone distribution in one-dimensional sub-monolayer film growth:
a fragmentation theory approach}
\author{M Grinfeld$^1$, W Lamb$^1$, K P O'Neill$^1$, and P A Mulheran$^2$}

\address{$^1$ Department of Mathematics and Statistics, University of
Strathclyde, 26 Richmond Street, Glasgow G1 1XH, UK}
\eads{\mailto{m.grinfeld@strath.ac.uk},\mailto{w.lamb@strath.ac.uk},
\mailto{kenneth.o-neill@strath.ac.uk}}

\address{$^2$ Department of Chemical and Process Engineering, University of
Strathclyde, 75 Montrose Street, Glasgow G1 1XJ,
UK}\ead{paul.mulheran@strath.ac.uk}

\begin{abstract}

\noindent The distribution of capture zones formed during the nucleation and
growth of point islands on a one-dimensional substrate during monomer
deposition is considered for general critical island size $i$. A
fragmentation theory approach yields the small and (for $i=0$) large size
asymptotics for the capture zone distribution (CZD) under the assumption of
no neighbour-neighbour gap size correlation. These CZD asymptotic forms are
different to those of the Generalised Wigner Surmise which has recently been
proposed for island nucleation and growth models, and we discuss the reasons
for the discrepancies.

\end{abstract}


\section{Introduction}
\label{sec1}

Island nucleation and growth during the early stages of thin film
deposition has attracted much attention over the last few decades
\cite{Ven,Brune,Mul09}. The islands that arise from the aggregation of
the deposited material form the building blocks for nanostructure
growth \cite{RR,Fanfoni} and subsequent film morphology \cite{Mul08},
and so are of direct technological interest. The growth process is
amenable to Monte Carlo (MC) simulation, which in turn have inspired many
theoretical analyses which revisit meanfield rate equations (for example
see references \cite{BE,AFL,RZ}). The most striking aspect to arise
from these models has been the identification of scale invariance in
the growth process, exemplified in the island size distribution (ISD)
\cite{AF}. It has been recognised that in order to understand the ISD,
one cannot simply rely on mean field equations since they do not always
yield the correct form for it \cite{B96} and in some cases do not even
display the fundamental scaling property seen in simulation and
experiment \cite{BC,SP}.

In order to better understand the ISD, we must consider the growth
rate of an island that has nucleated on the substrate. It grows by
capturing deposited monomers, which diffuse to it across the
substrate. The growth rate of the island is therefore dependent on its
capture zone, defined as the region of substrate closer to this island
than to any other \cite{MB,MR,EB}; monomers deposited into the
island's capture zone are more likely to diffuse to it than to any
other competing island. This concept has been successfully used for
the ISD found in growth on both two- \cite{MB,B96} and one-dimensional
substrates \cite{BM}. The latter model is of particular interest,
since it can be analysed in great detail, and it is the subject of the
current paper.

Recently there has been renewed interest in the capture zones and associated
size distribution (CZD) where it has been proposed that the CZD follows the
Generalised Wigner Surmise \cite{PE,PE10}. In particular, it has
been suggested \cite{PE} that the CZD obeys
\begin{equation}\label{GWS}
P(s) = a_\beta s^\beta \exp(-b_\beta s^2),
\end{equation}
where
\[
a_\beta = \frac{2\Gamma \left( \frac{\beta+2}{2}
\right)^{\beta+1}}{\Gamma \left( \frac{\beta+1}{2}
\right)^{\beta+2}}, \; \;
b_\beta = \left[\frac{\Gamma \left( \frac{\beta+2}{2}
\right)}{\Gamma \left( \frac{\beta+1}{2}
\right)}\right]^2
\]
are normalisation constants so that
\[
\int_0^\infty P(s) \, ds = \int_0^\infty sP(s) \, ds =1,
\]
and
\begin{equation}\label{eq:beta}
\beta=
\begin{cases}
\frac{2}{d}(i+1) & \mbox{ if } $d$=1,2  \\
(i+1) & \mbox{ if } $d$=3.
\end{cases}
\end{equation}
Here $ s=x/\langle x\rangle$ is the capture zone size $x$ scaled to its
average $\langle x\rangle$ at any time $t$ in the aggregation regime
\cite{AFL}. The aggregation regime is such that the island density is
greater than the monomer density, so that further nucleation is a slow
process compared to the island growth due to monomer capture.

Despite the excellent visual comparisons between equation (\ref{GWS})
and MC data taken from the literature \cite{PE}, the validity
of the GWS has already been challenged. Shi et al. \cite{SSA} studied
$i=1$ point island models in $d=1,2,3,4$ dimensions, finding that the
CZD is more sharply peaked than the GWS suggests for $d=1,4$ and a
better choice of $\beta=3$ rather than $\beta=2$ for $d=2,3$. Li et
al. \cite{LHE} also question the $i=1$, $d=2$ GWS form for the CZD,
proposing their own form based on a sophisticated theory for capture
zone evolution in two dimensions \cite{EB}.

We also note that there is a conflict in the small and large size form
of (\ref{GWS}) and that of the theory presented in \cite{BM} for the
one-dimensional (1-D) nucleation and growth of point islands. The work
in \cite{BM} focused on the case with critical island size $i=1$,
where we impose the condition that $i+1$ monomers must coincide in
order to nucleate a stable, immobile island. The analysis in \cite{BM}
utilises a fragmentation theory approach (as described below) which we
consider to be physically reasonable. It is therefore of interest to
ask whether a similar conflict exists for other critical island sizes.
In this work we extend the 1-D point-island fragmentation approach to 
$i=0,1,2,3,{\ldots}$ and show that the small-size conflict with
\cite{PE} holds for all non-negative integers $i$. The case of $i=1$
in this model has recently been considered in \cite{GPE}, where the
form of equation (\ref{GWS}) has been extended in light of MC
simulation data. Here we retain a focus on the comparison between the
original GWS of equation (\ref{GWS}) and the fragmentation model for
general $i$.

In \cite{BM} a MC simulation was presented, where monomers are
randomly deposited at rate $F$ onto a line of points which represents the
substrate. The monomers diffuse (with diffusion constant $D$) by nearest
neighbour hops until they nucleate a new island, or are adsorbed by hopping
onto an existing island. Since nucleation is a rare event in the aggregation
regime, it is assumed \cite{BM} that the average monomer density $n_1(x)$ in
the gap (length y) between two neighbouring islands obeys its saturated form
\begin{equation}\label{md}
n_1(x) = Rx(y-x),\; \;  0<x<y,
\end{equation}
where we have set $R={\displaystyle \frac{F}{2D}}$. 

Furthermore, this density profile is then used to predict the probability of
a nucleation occurring at position $x$ in the gap, which we take to be
proportional to $n_1^{i+1}(x)$. In adopting this form, it is assumed that
the nucleation arises from the congregation of $(i+1)$ ``mature'' monomers
that have explored their gap and lost memory of their deposition events.

In this work we will adopt the same approach as in \cite{BM} and
derive a gap-size evolution equation for $i=0,1,2,\ldots$. From this
equation, we obtain information on similarity solutions which involve
a reduced gap size distribution function $\phi(y)$. We are then able to
construct the CZD under the assumption used in \cite{BM} that due to
the stochastic nature of the nucleation process there are no
neighbour-neighbour gap size correlations in the system. Under this
assumption, we can write
\begin{equation}\label{eq:CZD}
P(s)= 2 \int_0^{2s} \phi(y) \phi(2s-y) \, dy.
\end{equation}

We shall then show that the small size form for $P(s)$ given by
equation (\ref{eq:CZD}) is different to that of equation (\ref{GWS}).
In particular, the small size power dependence from equation
(\ref{eq:CZD}) will prove to be an odd power of $s$, whereas equation
(\ref{GWS}) always gives an even power for one-dimensional
substrate. We also prove that the large-size forms in equations
(\ref{GWS}) and (\ref{eq:CZD}) for the spontaneous nucleation model
($i=0$) are in similar disagreement.

\section{The gap evolution equation}
\label{sec2} 

We model the one-dimensional aggregation regime by regarding nucleated islands as points on a line. Our aim
is to obtain an equation that describes the evolution of the gaps between adjacent islands. Let $n(x,t)$ denote
the number concentration of gaps of width $x$ at time $t$. Since any new nucleation event that occurs in a `parent' gap of
width $y$ will result in the creation of two `daughter' gaps of widths $x$ and $y-x$, it is clear that the
evolution of the system of gap sizes can be interpreted as a fragmentation process. Processes of this type are usually modelled by the linear fragmentation equation
\begin{equation}\label{1}
\frac{\partial}{\partial t}\,n(x,t) = -a(x)n(x,t) + \int_x^\infty b(x|y)a(y)n(y,t)\,dy.
 \end{equation}
In the case of gap sizes, $a(x)$ represents the rate at which gaps of width $x$ fragment (due to nucleation) and
$b(x|y)$ describes the distribution of gaps of width $x$ resulting from the fragmentation of a parent gap of width $y > x$. Since each new nucleation leads to the birth of only two daughter gaps, and since the total length of all the gaps is preserved, we require $b$ to satisfy
\begin{equation}\label{bconds}
\int_0^y\,b(x|y)\,dx = 2  \quad \mbox{ and } \quad
\int_0^y\,xb(x|y)\, dx = y.
\end{equation}
To obtain appropriate functions $a$ and $b$, as explained above, 
we follow the approach used by Blackman and Mulheran \cite{BM}. Therefore,
we assume that, in the aggregation regime, the density of monomers in a gap
of width $y$ is given by (\ref{md}). 

By the reasoning given in the Introduction, the probability of a new
nucleation occurring in a gap of width $y$ may be taken as being
proportional to
\begin{equation}\label{a}
  a(y) = R^{i+1}\int_0^y\,x^{i+1}(y-x)^{i+1}\,dx = R^{i+1}
  B(i+2,i+2)\,y^{2i+3},
\end{equation}
where $B(\cdot, \cdot)$ denotes the Beta function. Then, given that a
nucleation event has caused a gap of width $y$ to fragment, with
fragmentation rate $a(y)$, the probability that it will occur at a scaled
position $r=x/y$ in $[r_0,r_0+dr] \subseteq [0,1]$ is given by $h(r)dr$,
where
 \begin{equation}\label{h}
 h(r) = R^{i+1}r^{i+1}(1-r)^{i+1},\ 0 \leq r \leq 1.
 \end{equation}
  Since $x \in [x_0,x_0+dx] \Leftrightarrow x/y \in [x_0/y, x_0/y + dx/y] \Leftrightarrow r \in [r_0,r_0 + dr]$, the probability that the nucleation occurs at $x \in [x_0,x_0 + dx]$ is given by $h(x_0/y)dx/y$.
    Therefore, we take
 \begin{equation}\label{b}
 b(x|y) = k\,h(x/y)/y,
 \end{equation}
 where 
 \[
 k = \frac{1}{R^{i+1} B(i+3,i+2)}
 \]
 so that (\ref{bconds}) is satisfied. This leads to
  \[
 b(x|y) =  y^{-2i-3}x^{i+1}(y-x)^{i+1}/B(i+3,i+2),
 \]
and equation (\ref{1}) therefore becomes
\begin{eqnarray}
\frac{\partial}{\partial t}\,n(x,t) &&= R^{i+1} \left(-B(i+2,i+2)\,x^{2i+3}n(x,t)
  \phantom{+\int}\right.
\nonumber \\
&&\qquad \qquad+ 2\left.\int_x^\infty x^{i+1}(y-x)^{i+1}n(y,t)\,dy\right). \label{2}
 \end{eqnarray}
A simple re-scaling of the time variable then yields the gap-size evolution equation
\begin{equation}\label{3}
\frac{\partial}{\partial t}\,n(x,t) =  -B(i+2,i+2)\,x^{2i+3}n(x,t) +
2 \int_x^\infty x^{i+1}(y-x)^{i+1}n(y,t)\,dy.
 \end{equation}
This equation is a particular case of the linear, homogeneous fragmentation equation
\begin{equation}\label{lhe}
\frac{\partial}{\partial t}n(x,t) =  - c_\sigma x^\sigma n(x,t) +
c_\sigma \int_x^\infty y^{\sigma-1} K(x/y)n(y,t)\,dy
\end{equation}
for some $\sigma \geq 0$ in which $K(x/y)$, a homogeneous function of degree
zero, determines the number of daughter ``particles'' of size $x$ obtained
when a parent ``particle'' of size $y > x$ fragments.  There is a
considerable literature on equations of this type, and various mathematical
techniques have been used in the analysis of (\ref{lhe}). For example, the
theory of strongly continuous semigroups of operators can be applied to
establish the existence and uniqueness of physically meaningful solutions;
see, for example, \cite[Chapter 8]{BaAr}. Similarity solutions have also
been investigated by a number of authors, including Ziff and McGrady
\cite{ZM} (for the case $i=0$ in equation (\ref{3})), Cheng and Redner
\cite{CR} and Treat \cite{Treat}. As shown in \cite{Treat}, similarity
solutions can be written in the form
\begin{equation}\label{simsol}
n^*(x,t) = \frac{N^{*2}(t)}{V}\, \phi\left(\frac{N^*(t)x}{V}\right)
\end{equation}
where the \emph{reduced distribution} $\phi$ is required to satisfy an integral equation and is normalised so that
\[
\int_0^\infty \phi(y)\,dy = \int_0^\infty y\phi(y)\,dy = 1,
\]
and
\[
N^*(t) := \int_0^\infty n^*(x,t)\,dx, \quad V:= \int_0^\infty xn^*(x,t)\,dx,
\]
are, respectively,  the zeroth and first moments of $n^*$. An explicit expression, involving the Meijer $G$-function, is derived in \cite[Section 6]{Treat} for the specific case when the \emph{daughter distribution} function $K$ in equation (\ref{lhe}) takes the form
\begin{equation}\label{spec}
K(r) = r^\gamma(b_0 + b_1r + \cdots + b_pr^p), 
\end{equation}
$p$ is a non-negative integer, $b_0,\, \ldots , b_p \in \R$, and $0
\leq r \leq 1$. We shall make use of the simple case $p=1$, discussed
in \cite[Section 7.2]{Treat}, in the next section. 

Asymptotic properties of the function $\phi$ have also been established for
more general homogeneous functions $K$. In particular, in \cite{CR} and
\cite[Section 5]{Treat}, it is shown that
\begin{equation}
 \phi(y) = {\cal O}(y^\gamma) \mbox{ as } y \to 0,  \label{as1}\\
\end{equation}
provided that 
$\displaystyle \lim_{r\rightarrow 0} r^{-\gamma-2}\int_0^r sK(s)ds$ 
exists and is non-zero, and

\begin{equation}
 \phi(y) = {\cal O} y^{K(1)-2}\exp(-c y^{\sigma})\  \mbox{ as } \ y \to \infty, \label{as2}
\end{equation}
for some constant $c>0$.

The questions of existence and stability, in an appropriately defined sense,
of similarity solutions to fragmentation equations of homogeneous type have
also been addressed in \cite{EMR}.

In the case of the gap fragmentation equation (\ref{3}), (\ref{as1}) and
(\ref{as2}) lead directly to the following result.

\begin{theorem}\label{theo:CR} The equation (\ref{3}) has a similarity solution of the form (\ref{simsol}) where the gap size distribution $\phi$ satisfies
\begin{description}
\item{\rm I.}\  $\phi(y) = {\cal O}(y^{i+1})$ as $y \rightarrow 0$;
\item{\rm II.}\  $\phi(y) = {\cal O}(y^{-2}\exp(-c y^{2i+3}))$ as $y \rightarrow
\infty$ for some $c>0$.
\end{description}
\end{theorem}

The task now is to understand, given the information we have for
$\phi(y)$, the behaviour of the convolution (\ref{eq:CZD}). The
situation for small $s$ is straightforward for any $i$, but for large
$s$ we only have a result if $i=0$. The derivation of the asymptotics
is non-trivial and requires the use of a multi-dimensional version of
Laplace's method. This is the content of the next two sections.

\section{CZD asymptotics}

The small $s$ behaviour of $P(s)$ for arbitrary $i$ can be obtained
immediately from part I of Theorem \ref{theo:CR} and equation
(\ref{eq:CZD}):

\begin{theorem}\label{smalls}
For the critical island size $i \ge 0$, we have
\[
P(s) ={\cal O} (s^{2i+3}) \hbox{ as } s \rightarrow 0,
\]
where $P(s)$ is the 1-D CZD.
\end{theorem} 
\proof \ We have, for small $s$ and $0 < y < 2s$,
\begin{displaymath}
 \phi(y) = {\cal O} (y^{i+1}), \ \phi(2s-y) = {\cal O}((2s-y)^{i+1}),
\end{displaymath}
Hence $\phi(y)\phi(2s-y)={\cal O}(s^{i+1}y^{i+1})$
and therefore
\[
 P(s) = 2\int_0^{2s} \phi(y)\phi(2s-y)\, dy =
{\cal O}(s^{i+1}s^{i+2}) = {\cal O}(s^{2i+3})
\]
as $s \rightarrow 0$.
\qed
 
By Theorem \ref{smalls}, the exponent is always odd which differs from
the GWS prediction that we should have $P(s) ={\cal O}(s^\beta)$ where
$\beta = 2(i+1)$ is always even when $d =1 $. Hence we are led to the
conclusion that the GWS does not describe the behaviour of the scaling
function $P(s)$ as $s \rightarrow 0$ for any $i \geq 0$ if we accept
that the gap evolution equation (\ref{3}), and the relation between
$\phi(y)$ and the capture zone distribution $P(s)$ given by
(\ref{eq:CZD}), are correct. 

The next aim is to understand the asymptotic behaviour of $P(s)$ in 
(\ref{eq:CZD}) for large $s$. For this, it would appear that an explicit
and reasonably simple, formula for the reduced distribution $\phi(y)$ is
required. In the case $i =0$, the gap fragmentation equation (\ref{3})
becomes 
\[ 
\frac{\partial}{\partial t} n(x,t) = -\frac{x^3}{6}n(x,t) +
2\int_x^\infty x(y-x)n(y,t)\,dy, 
\] 
which is the equation analysed by Ziff and McGrady in \cite{ZM}. Note
also, that in this case the homogeneous function $K$ can be obtained
from the general \emph{linear daughter distribution}, 
\begin{equation}\label{speclin}
K(r) = r^\gamma(b_0 + b_1r), 
\end{equation}
investigated by Treat in \cite[Section 7.2]{Treat}, by setting $\gamma = 1,
b_0 = 12, b_1 = -12.$ On applying \cite[(7.6)]{Treat}, we deduce
that $\phi(y) = \overline{\phi}(\eta)$, where $\eta = y^3/\mu^3,$
\[
\mu = \frac{\Gamma(2/3)\Gamma(7/3)}{\Gamma(4/3)\Gamma(2)} = \frac{4}{3}\Gamma(2/3)
\]
and
\begin{eqnarray*}
\overline{\phi} (\eta) &=&\frac{A}{\mu}\eta^{1/3}e^{-\eta} 
\frac{1}{\Gamma(1)} \int_0^\infty (1+s)^{-4/3}\exp(-s\eta)\,ds, 
\quad A = 3/\Gamma(2/3),\\
&=& \frac{A}{\mu}\eta^{1/3}\int_1^\infty z^{-4/3} \exp(-z\eta) \,dz
= \frac{A}{\mu} \eta^{2/3}\int_\eta^\infty u^{-4/3}\exp(-u)\,du.
\end{eqnarray*}
Hence
\begin{eqnarray}\label{eqLAP2}
\phi(y) & = & \frac{A}{\mu^3}y^2\int_{y^3/\mu^3}^\infty u^{-4/3} \exp(-u)\,du \\
\label{eqLAP2.1}
& = & \frac{3y}{\Gamma(\frac{2}{3}) \mu^2} \int_1^{\infty} v^{-4/3}e^{-v(y/\mu)^3}dv.
\end{eqnarray}
Note that Ziff and McGrady proposed a similar formula for $\phi$ but with
the constant $\mu^3 (\approx 5.88)$ in the lower limit of integration in (\ref{eqLAP2})
replaced by $6$. That the correct choice is $\mu^3$ follows from the fact
that this leads to 
\[
\int_0^\infty \phi(y)\,dy = \int_0^\infty y \phi(y)\,dy = 1.
\]

This explicit representation of $\phi(y)$ for the case $i =0$ enables us to
obtain the following result.
\begin{theorem}\label{thm:ZM}
If $i=0$ then
\[
P(s) ={\cal O}\left( s^{-9/2} \exp \left( -2\frac{s^3}{\mu^3} \right) \right)
\hbox{ as } s \rightarrow \infty.
\]
\end{theorem}

The next section is devoted to proving this theorem.

\section{Proof of Theorem \ref{thm:ZM}}

The arguments of Wong \cite[Ch. VIII, Sections 8 and 11]{Wong} allow us to
obtain the following general theorem about two-dimensional Laplace
integrals. 

Let
\begin{equation}
 \label{eqLAP5}
 J(\lambda) = \int \! \! \! \int_D g(v,w)e^{-\lambda f(v,w)} dvdw,
\end{equation}
where $\lambda$ is a large parameter,
$D \subset \mathbb{R}^2$, and let $f(v,w)$, $g(v,w)$ be smooth functions on 
$D$. 

We have
\begin{theorem}\label{th:Lap}
If $(v_0,w_0)$ is the global minimum of $f(v,w)$ on $\overline{D}$ and is a
critical point of the third kind, then as $\lambda \rightarrow \infty$, 
$J(\lambda) ={\cal O}( \lambda^{-2}e^{-\lambda f(v_0,w_0)})$. 
\end{theorem}

We remind the reader that $(v_0,w_0)$ is a \textbf{critical point of the
third kind} of $f(v,w)$ if it is an extremum point of $f$ on $D$
belonging to the boundary of $D$ through which pass two intersecting
tangent lines, neither of which coincides with the tangent to the
level curve $\psi(v,w)=\psi(v_0,w_0)$ at $(v_0,w_0)$.

Our aim is to understand the asymptotics of (\ref{eq:CZD}) as $s \rightarrow
\infty$ in the case of $i=0$ using the explicit form (\ref{eqLAP2.1}).

If we let $u=y^3v/\mu^3$ in  (\ref{eqLAP2}), substitute the resulting
expression for $\phi(y)$ into (\ref{eq:CZD}) and then put $y=2sz$, we have
\begin{equation}
\label{eqLAP3}
 P(s) = \frac{144s^3}{\mu^4 \Gamma(2/3)^2} 
 \int_1^{\infty} \int_1^{\infty} (vw)^{-4/3} \int_0^1
 z(1-z)e^{-(2s)^3(z^3v+(1-z)^3 w)/\mu^3} dzdvdw.
\end{equation}

Our strategy is to use Laplace's method to evaluate the inner integral,
which will give us a two-dimensional Laplace integral on $D=(1,\infty)
\times (1,\infty)$, which will be attacked using Theorem \ref{th:Lap}.

For any $(v,w) \in D$, set $S(z)=z^3v+(1-z)^3w$. If $v=w$, $S(z)$ has a
unique minimum at $z_+=1/2$. If $v \neq w$, critical points of $S(z)$ satisfy
$(v-w)z^2+2wz-w = 0$, so that
\[
z_{\pm} = \frac{-w \pm \sqrt{vw}}{v-w}.
\]
It is easy to check that $S''(z_+) = 6\sqrt{vw} >0$, so that the minimum of
$S(z)$ is obtained at
\begin{equation}
 \label{eqLAP3.1}
 z_+= \frac{-w + \sqrt{vw}}{v-w} = \frac{\sqrt{w}}{\sqrt{w}+\sqrt{v}}
\in (0,1).
\end{equation}

Hence by Laplace's formula, setting ${\displaystyle \lambda= \frac{8
s^3}{\mu^3}}$,
we have, as $\lambda \rightarrow \infty$,
\[
I_1(\lambda) := \int_0^1 z(1-z)\exp(-\lambda S(z))\, dz \sim
z_+(1-z_+)e^{-\lambda S(z_+)} \sqrt{\frac{2\pi}{\lambda S''(z_+)}}.
\]
Note that
\[
S(z_+)= \frac{wv}{(\sqrt{w}+\sqrt{v})^2},
\]
which reaches its minimum value $1/4$ on $D$ at the corner point
$(1,1)$.

Thus, the inner integral in (\ref{eqLAP3}) satisfies, as $s \rightarrow \infty$,
\[
I_1\left( \frac{8s^3}{\mu^3} \right) \sim \left(\frac{8s^3}{\mu^3}\right)^{-1/2}\rho(v,w) \exp \left(
-\frac{8s^3}{\mu^3}
\frac{wv}{(\sqrt{w}+\sqrt{v})^2} \right),
\]
where we have put
\[
\rho(v,w)= \sqrt{\frac{2\pi}{S''(z_+)}}z_+(1-z_+).
\]

This means by (\ref{eqLAP3}) that
\begin{equation}
 \label{eqLAP4}
P(s) ={\cal O}\left( s^{3/2} \int_1^{\infty} \int_1^{\infty} \rho(v,w) (vw)^{-4/3}
\exp \left( -\frac{8s^3}{\mu^3}
\frac{wv}{(\sqrt{w}+\sqrt{v})^2} \right) dvdw \right),
\end{equation}
and now invoking Theorem \ref{th:Lap} we have, as $s \rightarrow \infty$,
\[
P(s) ={\cal O}\left( s^{3/2}e^{-(8s^3/\mu^3)(1/4)} \left( \frac{8s^3}{\mu^3} \right)^{-2}\right),
\]
i.e. $P(s) = {\cal O} \left( s^{-9/2}e^{-2s^3/\mu^3} \right)$ as required.

\section{Conclusions}


In this paper we have shown that the Blackman and Mulheran model \cite{BM}
for the nucleation and growth of point islands on a one-dimensional (1-D)
substrate, based on a fragmentation equation approach, yields a capture zone distribution (CZD) that is
different to the Generalised Wigner Surmise (GWS) in equation (\ref{GWS}) \cite{PE}.
The asymptotics of the CZD are measurable in MC simulation,
allowing these theories to be tested, and we will present our analyses of
simulations elsewhere \cite{OGLM}. Here we will conclude this current work
with a discussion of how the two theoretical approaches differ and how they
might be reconciled.

It is interesting to note that the justification for the relationship
between the parameter $\beta$ in equation (\ref{GWS}) and the critical island
size, as given in equation (\ref{eq:beta}), is based on the same physical model
analysed in this paper. In \cite{PE}, the island nucleation rate is
discussed in terms of the monomer density n, and the probability of $(i+1)$
monomers coinciding is used to give the nucleation rate as $\sim n^{i+1}$.
This of course is exactly the physical basis we have used, so it is
worthwhile considering why \cite{PE} end up with different power-law
behaviour for small capture zone sizes.

Let us summarise the phenomenological arguments used in \cite{PE}.
Firstly the authors consider the nucleation rate within capture zones
with size $s$, which have a relative density $P(s)$, using the mean
field value of the monomer density $n$. They then reconsider the
nucleation rate within a small zone of size $s$ using the locally
averaged monomer density. In one space dimension, they argue that this
density $ \sim n s^{2}$. Equating the two rates, $P(s) n^ {i+1} \sim
(n \, s^{2})^{i+1}$ yields their law for $\beta$ in 1-D nucleation so
that $P(s)\sim s^{2i+2}$ for small $s$. We note that recently the
originators of equation (\ref{GWS}) have considered alternative forms
to the Gaussian tail following evidence from their own MC
simulations with $i=1$.

From our perspective, we see a number of problems with this argument.
Firstly, the method of constructing the two alternative nucleation rates
appears internally inconsistent, using both the mean field monomer density
$n$ and a local approximation. Usually this approach would involve taking
averages of the latter to reach consistency with the former \cite{BC, BM},
however this was not done in \cite{PE}. The second criticism, more pertinent
to this paper, is that the nucleation rate using the locally averaged monomer
density within a capture zone has not been justified. Indeed, in
\cite{PE10}, the authors revise their argument for two dimensional substrates and use a
spatially dependent monomer density $n(r)$ within a capture zone to derive
the nucleation rate for the zone. This involves integrating $n^{i+1}(r)$
over the zone, reminiscent of the approach used elsewhere for the two
dimensional substrate \cite{PAM04}, resulting in a different relationship
$\beta(i)$ which better fits Monte Carlo data.\footnote{In \cite{PE10}, the
authors use $n(r) \sim R^2 - r^2$, $R_i < r < R$, with $R_i$ and $R$ being
the island and circular capture zone radii. It is not clear what motivated
the choice of boundary conditions for this form; we would have expected the
monomer density to be zero at the island edge, rising to a maximum at the
capture zone radius so that its gradient is zero at this boundary
\cite{PAM04}.}

Presumably, this type of modification can be taken forward to the 1-D case
discussed here. Following the same line of argument, we might take the local
monomer density in a small capture zone size $s$ as that from two gaps of
size $s/2$, using the form of equation (\ref{md}) above. If we do this, we
will end up with the formula $\beta(i) = 2i+3$, thereby agreeing with the
small-size CZD behaviour we have derived in this work. Although this does go
some way to reconciling the two approaches, we consider that this line of
reasoning is heuristic at best, and much prefer the more rigorous methods
adopted in our work presented here in this paper.

The second component of the GWS in equation (\ref{GWS}) is the
large-size behaviour of the CZD. Here too we see a conflict with the
fragmentation theory approach we adopt, again despite the same physical
basis for the models. In \cite{PE}, the authors are motivated by the idea of
a large fluctuating capture zone being constrained by its neighbours, rather
than any attempt to consider in detail how the nucleation process impacts on
the large size CZD behaviour. In this work we have been able to prove that
for $i=0$, the large size dependence of the CZD following equation
(\ref{eq:CZD}) mirrors that of the corresponding gap distribution $\phi(y)$.
In particular, we have $P(s) \sim \exp(-2 s^{3}/\mu^3)$ rather than the Gaussian
tail of equation (\ref{GWS}). Furthermore, whilst it remains unproved, we can
conjecture that the CZD for larger values of $i$ will similarly follow the
asymptotics of the corresponding GSD and follow $\exp(-c_i s^{2i+3})$ for
some constants $c_i$ in stark contrast to the Gaussian tail.

In conclusion, we have shown that the GWS of \cite{PE} is in conflict
with the asymptotic solutions to the fragmentation theory analysis of
point island nucleation and growth in one dimension. Given that the
physical basis of the two approaches is the same, we believe that
these differences show important failings of the GWS. Of course, this
is not to claim that the Blackman and Mulheran model, embodied in
equations (\ref{md}) and (\ref{eq:CZD}) above, has been proven to be
correct; confrontation with simulation and experiment will ultimately
arbitrate between these theories. Along these lines, we note the
recent analysis of the case of $i=1$ in \cite{GPE}, and our own
detailed comparisons with extensive simulation data for $i=0,1,2,3$
presented in \cite{OGLM}.


\begin{thebibliography}{99}

\bibitem{AF}  Amar J G and Family F 1995 {\it Phys. Rev. Lett.} {\bf 74} 2066--9 

\bibitem{AFL} Amar J G , Family F, and Lam P-M  1994 {\it Phys. Rev.} B {\bf
50}  8781--97 

\bibitem{BC}  Bales G S  and Chrzan D C  1994 {\it Phys. Rev.} B {\bf 50}
6057--67


\bibitem{BaAr} Banasiak J and Arlotti L 2006  {\it Positive Perturbations of
Semigroups with Applications} (London: Springer Verlag) p 203



\bibitem{BE} Bartelt M C and Evans J W 1992 {\it Phys. Rev.} B {\bf 46} 
12675--87

\bibitem{B96} Bartelt M C  and Evans J W 1996  {\it Phys. Rev.} B {\bf
54 } R17359--62


\bibitem{BM} Blackman J A and Mulheran P A 1996 {\it Phys. Rev.} B {\bf
54} 11681--92

\bibitem{Brune} Brune H 1998 {\it Surf. Sci. Rep.} {\bf 31} 125--229

\bibitem{CR} Cheng Z and Redner S 1988 {\it Phys. Rev. Lett.} {\bf 60} 2450--3

\bibitem{EMR} Escobedo M, Mischler S, and Rodriguez Ricard M
2005 {\it Ann. H. Poincar\'e Anal. Nonlin.} {\bf 22} 99--125

\bibitem{EB} Evans J W and Bartelt M C 2002 {\it Phys. Rev.\/} B {\bf 66}
235410--21 


\bibitem{Fanfoni} Fanfoni M 2008 {\it  J. Phys. Condens. Matter} {\bf 20}
015222

\bibitem{GPE} Gonzalez D L, Pimpinelli A, and Einstein T L 2011 {\it
Phys. Rev. } E {\bf 84} 011601--12

\bibitem{LHE}   Li M, Han Y, and Evans J W 2010 {\it Phys. Rev. Lett.\/}
{\bf 104} 149601

\bibitem{PAM04}  Mulheran P A 2004 {\it Europhys. Lett.\/} {\bf 65} 379--85

\bibitem{Mul08} Mulheran P A, Pellenc D, Bennett R A, Green R J, and Sperrin
M  2008 {\it Phys. Rev. Lett.} {\bf 100} 068102

\bibitem{Mul09} Mulheran P A 2009 Theory of cluster growth on surfaces {\it Metallic Nanoparticles\/} ({\it
Handbook of Metal Physics\/} vol. 5) ed J A Blackman (Amsterdam: Elsevier)
pp 73--111

\bibitem{MB} Mulheran P A and Blackman J A  1996 {\it Phys. Rev.\/} B {\bf  
53} 10261--7

\bibitem{MR} Mulheran P A and Robbie D A 2000 {\it Europhys. Lett.\/} {\bf 49}
617--23

\bibitem{OGLM} O'Neill K P, Grinfeld M, Lamb W, and Mulheran P A 2011 Gap
size and capture zone distributions in one-dimensional point island 
nucleation and growth simulations: asymptotics and models, e-print
arXiv:1110.2332 (submitted to {\it Phys. Rev. } E)


\bibitem{PE} Pimpinelli A and Einstein T L 2007 {\it Phys. Rev. Lett.}
{\bf 99} 226102

\bibitem{PE10}  Pimpinelli A and Einstein T L 2010 {\it Phys. Rev. Lett.}
{\bf 104} 149602


\bibitem{RZ} Ratsch C, Zangwill A, Smilauer P,  and Vvedensky D D 1994 {\it Phys.
Rev. Lett.} {\bf 72} 3194--7 


\bibitem{RR} Ratto R and Rosei F 2010 {\it Mater. Sci. Eng. Rep.} {\bf 70}
243--64

\bibitem{SSA} Shi F, Shim Y, and Amar J G 2009 {\it Phys. Rev.\/} B 011602

\bibitem{SP} Stroscio J A and Pierce D T 1994 {\it Phys. Rev.\/} B {\bf 49}  
8522--5

\bibitem{Treat} Treat R P 1997  {\it J. Phys.} A: {\it Math. Gen.} {\bf 30} 2519--43

\bibitem{Ven} Venables J A 1973 {\it Philos. Mag.}  {\bf 27}  693--738

\bibitem{Wong} R. Wong, {\em Asymptotic Approximations of Integrals},
(Philadelphia: SIAM) 2001 pp 448, 459


\bibitem{ZM} Ziff R M and McGrady  E D 1986  {\it Macromolecules} {\bf 19} 
2513--19


\end{thebibliography}
\end{document}